\documentclass{PoS}
\rightline{\parbox{2cm}{\large\rm
RBRC-616
}}

\usepackage{amssymb}
\usepackage{amsmath}

\newcommand{\be}{\begin{equation}}
\newcommand{\ee}{\end{equation}}
\newcommand{\bea}{\begin{eqnarray}}
\newcommand{\eea}{\end{eqnarray}}
\newcommand{\bi}{\begin{itemize}}
\newcommand{\ei}{\end{itemize}}
\newcommand{\bc}{\begin{center}}
\newcommand{\ec}{\end{center}}

\newcommand{\tr}{{\rm Tr}}

\title{$SU(3)$ breaking effects in hyperon beta decay from lattice QCD}

\ShortTitle{$SU(3)$ breaking effects in hyperon beta decay from lattice QCD}

\author{\speaker{Shoichi Sasaki}\thanks{We thanks to RIKEN, Brookhaven National Laboratory and the U.S. DOE for providing the facilities essential for the completion of this work.}\\
        RIKEN BNL Research Center, Brookhaven National Laboratory, Upton, NY 11973, USA and\\
        Department of Physics, University of Tokyo, Hongo, Tokyo 113-0033, JAPAN\\
        E-mail: \email{ssasaki@phys.s.u-tokyo.ac.jp}}

\author{Takeshi Yamazaki\thanks{
Present address: Physics Department, University of Connecticut, Storrs, CT 06269-3046, USA}\\
        RIKEN BNL Research Center, Brookhaven National Laboratory, Upton, NY 11973, USA\\
        E-mail: \email{yamazaki@quark.phy.bnl.gov}}

\abstract{We present results of an exploratory study of 
flavor $SU(3)$ breaking effects in hyperon beta decays
using domain wall fermions. From phenomenological point of view, 
the significance of this subject is twofold: (1) 
to extract the element $V_{us}$ of the Cabibbo-Kabayashi-Maskawa 
mixing matrix from the $\Delta S=1$ decay process, 
and (2) to provide vital information to analysis of 
the strange quark fraction of the proton spin with
the polarized deep inelastic scattering data.
In this study, we explore the $\Xi^0 \rightarrow \Sigma^+$ beta decay, 
which is highly sensitive to the $SU(3)$ breaking since this decay corresponds 
to the direct analogue of neutron beta decay under an exchange between the 
down quark and the strange quark. 
We expose the $SU(3)$ breaking effect on $g_A/g_V=g_1(0)/f_1(0)$
up to the first order in breaking. The second-class form factors
$g_2$ and $f_3$, of which non-zero values are the 
direct signals of the $SU(3)$ breaking effect, are also measured. 
Finally, we estimate $f_1(0)$ up to the second-order correction and then
evaluate $|V_{us}|$ combined with the KTeV experiment.
}

\FullConference{XXIV International Symposium on Lattice Field Theory\\
		 July 23-28 2006\\
		 Tucson Arizona, US}

\begin{document}

\section{Introduction}

The octet baryons ($p,n,\Lambda,\Sigma,\Xi$) admit various semileptonic decays (beta decays):
%
%
\be
B_1 \rightarrow B_2 + e^{-} + {\bar \nu}_e
\ee
which are described by the vector and axial-vector transitions
%
%
\be
\langle B_2 | V_{\alpha}(x)+A_{\alpha}(x) | B_1 \rangle 
={\overline u}_{B_2}(p') 
\left(
O^{V}_{\alpha}(q)+O^{A}_{\alpha}(q) 
\right)
u_{B_1}(p) e^{-iq\cdot x},
\ee
where $q\equiv p-p'$. Six form factors are needed to describe the hyperon beta decays:
the vector, weak magnetism and induced scalar form factors for
the vector current,
%
%
\be
O^{V}_{\alpha}(q) 
= \gamma_{\alpha} f_1(q^2) + \sigma_{\alpha \beta}q_{\beta} \frac{f_2(q^2)}{M_{B_1}}
- iq_{\alpha}\frac{f_3(q^2)}{M_{B_1}}
\label{Eq:VcMat}
\ee
and the axial-vector, weak electricity and induced pseudo-scalar from factors 
for the axial current,
%
%
\be
O^{A}_{\alpha}(q)
= \gamma_{\alpha}\gamma_5 g_1(q^2) 
+ \sigma_{\alpha \beta}q_{\beta}\gamma_5\frac{g_2(q^2)}{M_{B_1}} 
-iq_{\alpha} \gamma_5 \frac{g_3(q^2)}{M_{B_1}},
\label{Eq:AxMat}
\ee
which are here given in Euclidean space by performing
the Wick rotation from Minkowski space with $\gamma$-matrix conventions 
adopted in Refs.~\cite{{Cabibbo:2003cu},{Alavi-Harati:2001xk}}.
The forward limits of the vector and the axial-vector form factors are
well known as $g_V=f_1(0)$ and $g_A=g_1(0)$.
Form factors $f_3$ and $g_2$ are known as the second-class form factors,
which are identically zero in the certain symmetric limit (iso-spin symmetry,
$U$-spin symmetry or $V$-spin symmetry as $SU(2)$ subgroups of the 
flavor $SU(3)$ symmetry). 
For an example, the second-class form factors 
in neutron beta decay are prohibited to be non-zero values because of $G$-parity 
conservation in the iso-spin symmetry limit~\cite{Cabibbo:2003cu}.

From phenomenological point of view, the various ratios $g_A/g_V$ in the hyperon 
beta decays provide vital information to analysis of strange quark spin fraction 
inside the proton~\cite{Filippone:2001ux}. However, the flavor $SU(3)$ breaking 
effects are simply neglected in such analysis~\cite{Filippone:2001ux}, while sizeable 
effects are to be expected in general. 
Although there are various hyperon beta decays, we are especially interested in 
the $\Xi^0 \rightarrow \Sigma^+$ beta decay which is the direct analogue 
of neutron beta decay under an exchange between the down quark and the 
strange quark. If the flavor $SU(3)$ symmetry is manifest, the ratio $g_A/g_V$ 
should be identical to that of neutron beta decay. In other words, 
this particular decay is highly sensitive to the $SU(3)$ breaking. 
Indeed, some model (center-of-mass correction approach~\cite{Ratcliffe:1998su}) 
and the $1/N_c$ expansion approach~\cite{Flores-Mendieta:1998ii} 
predict that the $(g_A/g_V)_{\Xi\Sigma}$ is smaller than 
the $(g_A/g_V)_{np}$ by 8-10\% and 20-30\% respectively.
However, the first and single experiment done by the KTeV collaboration at FNAL
showed no indication of the $SU(3)$ breaking effect on 
$g_1(0)/f_1(0)=g_A/g_V$~\cite{Alavi-Harati:2001xk}.
The KTeV experiment also reported no evidence for a non-zero second-class form 
factor $g_2$~\cite{Alavi-Harati:2001xk}.

The $\Delta S=1$ beta decay processes such as the $\Xi^0 \rightarrow \Sigma^+$ 
beta decay are also applicable to determine
the element $V_{us}$ of the Cabibbo-Kabayashi-Maskawa mixing matrix 
other than the $K_{l3}$ decays~\cite{Leutwyler:1984je}. 
Experimentally, the product $|V_{us}f_1(0)|^2(1+3\left|g_1(0)/f_1(0)\right|^2)$
is obtained from the decay branching ratio and the value of
$g_1(0)/f_1(0)$ is determined by the measured asymmetries.
A theoretical input of $f_1(0)$, which is protected by the Ademollo-Gatto theorem
against the first-order correction in breaking~\cite{Ademollo:1964sr}, 
is necessary to evaluate $|V_{us}|$~\cite{Leutwyler:1984je}. 
Although theoretical accurate estimate of $f_1(0)$ 
are highly required for the precise determination of $|V_{us}|$,  
even a sign of the second-order correction, 
is somewhat controversial at present. 
The quark-model calculations show $f_1(0)/(f_1(0))_{SU(3)}<1$, while heavy baryon
chiral perturbation theory (HBChPT) and  large 
$N_c$ analysis, both predict $f_1(0)/(f_1(0))_{SU(3)}>1$~\cite{Mateu:2005wi}. 

In this study, we try to expose the flavor $SU(3)$ breaking effects
in the hyperon beta decay from the theoretical first principle calculation.

\section{Computational methodology}

First of all, we define the finite-momentum three-point functions 
for the relevant components of either the local vector current
 ($\Gamma_{\alpha}=V_{\alpha}$) or
the local axial current ($\Gamma_{\alpha}=A_{\alpha}$) with the interpolating
operators ${\cal B}_{1}$ and ${\cal B}_{2}$ for the $B_1$ and $B_2$ states:
\be
\langle {\cal B}_2(t', {\bf p '}){\Gamma}_{\alpha}(t, {\bf q}) \overline{\cal B}_1(0, {\bf p})\rangle
=
G_{\alpha}^{\Gamma}(p, p')\times f(t, t', E_{B_1}(p), E_{B_2}(p')) + \cdot\cdot\cdot,
\ee
where the initial ($B_1$) and final ($B_2$) states carry fixed momenta
${\bf p}$ and ${\bf p}'$ respectively and then the current operator
has a three-dimensional momentum transfer ${\bf q}={\bf p}-{\bf p}'$.
The ellipsis denotes excited state contributions which can be ignored in 
the case of $t'-t\gg 1$ and $t \gg 1$.
We separate the correlation function into two parts: $G_{\alpha}^{\Gamma}(p,p')$
which is defined as
%
%
\be
G^{\Gamma}_{\alpha}(p,p')=
(-i\gamma\cdot p'+M_{B_2})O_{\alpha}^{\Gamma}(q)(-i\gamma\cdot p+M_{B_1}),
\ee
where $O_{\alpha}^{\Gamma}(q)$ corresponds to either Eq.~(\ref{Eq:VcMat}) or
Eq.~(\ref{Eq:AxMat}),
and  the factor $f(t, t', E_{B_1}(p), E_{B_2}(p'))$ which collects all the kinematical
factors, normalization of states, and time dependence of the correlation function.

In this study, the semileptonic hyperon decay process $B_1 \rightarrow B_2$ is measured at the 
rest frame of the $B_2$ state ($|{\bf p}'|^2=0$), 
which leads to ${\bf q}={\bf p}$. Therefore,
the squared four-dimensional momentum transfer is given by 
$q^2=2M_{B_2}(E_{B_1}(p)-M_{B_1})-\Delta M^2$
where $\Delta M=M_{B_1}-M_{B_2}$ and $E_{B_1}(p)=\sqrt{M_{B_1}+{\bf p}^2}$. In the $SU(3)$ limit ($\Delta M=0$),
$q^2$ is always positive (space-like momentum). 
To extract the desired form factors, we take the trace of $G_{\alpha}^{\Gamma}(p,p'=0)$
with some appropriate projection operator ${\cal P}$ and then $\tr \{{\cal P} G_{\alpha}^{\Gamma}(p,p'=0)\}$ yields some linear combination of form factors
in each $\Gamma$ channel. We solve the simultaneous linear equations
in terms of form factors at fixed ${\bf q}$.

\section{Numerical results}

We have performed quenched lattice calculations on a $L^3 \times T
=16^3\times 32$ lattice with a renormalization group 
improved gauge action, DBW2 at $\beta=6/g^2=0.87$ ($a\approx 0.15$ fm).
Quark propagators were generated with three lighter quark masses $m_{ud}=$
0.04, 0.05 and 0.06 for up and down quarks and with two heavier quark masses 
$m_s=$0.08 and 0.10 for the strange quark, using domain wall fermions (DWFs) 
with $L_s=16$ and $M_5=1.8$. We refer to Ref.~\cite{Aoki:2002vt} for details of 
the quench DWF/DBW2 simulation.
In this study, we take 5 different combinations between
the up (down) quark and the strange quark as $(m_{ud}, m_s)$=(0.04, 0.08),
(0.05, 0.08), (0.06, 0.08), (0.04, 0.10) and (0.05, 0.10), which yield different
$SU(3)$ breaking patterns characterized by 
$\delta=(M_{B_1}-M_{B_2})/M_{B_1}=\Delta M/M_{B_1}$
in the range of 0.019 to 0.055.

Here we remark that the previous study of neutron beta decay 
with the same simulation parameters successfully yields a value of $g_A/g_V$
as $1.212\pm0.027$, which just underestimates the experimental one 
by less than 5\%~\cite{Sasaki:2003jh}.
This success encourages us to study the $SU(3)$ breaking effects in hyperon beta decay
through comparison between neutron beta decay and 
the $\Xi^0 \rightarrow \Sigma^+$ beta decay.

\subsection{First-order breaking in $g_1(0)/f_1(0)$}

\begin{figure}
\begin{minipage}[t]{0.47\linewidth}
\bc
\includegraphics[width=1.0\textwidth]{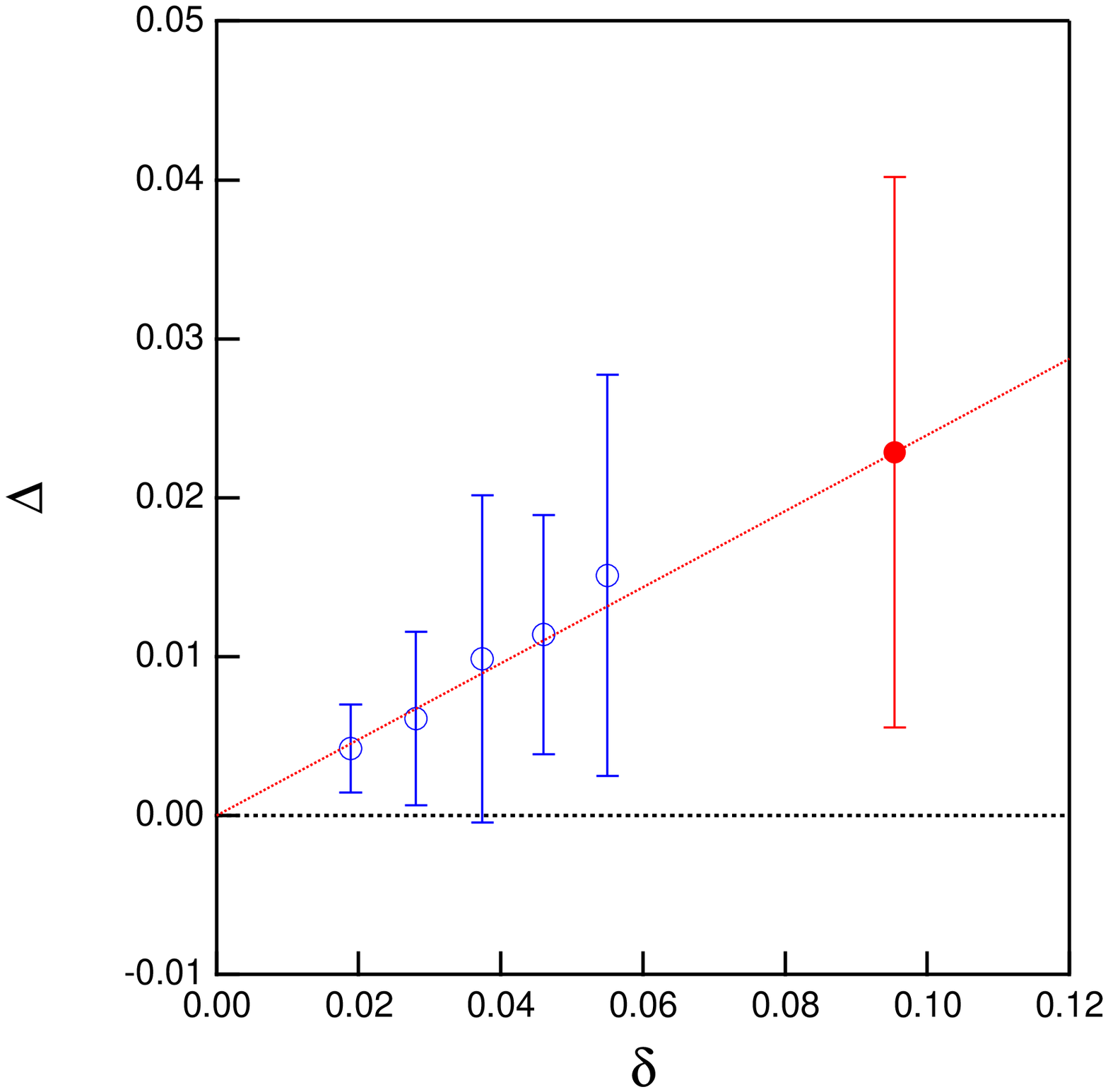}
\caption{The deviation from unity in the double ratio $R(t)$ as a function of the $SU(3)$ breaking parameter $\delta=\frac{M_{\Xi}-M_{\Sigma}}{M_{\Xi}}$.}
\label{FIG:g1f1ratio}
\ec
\end{minipage}
 \hspace*{0.05\linewidth}
\begin{minipage}[t]{0.47\linewidth}
\bc
\includegraphics[width=1.0\textwidth]{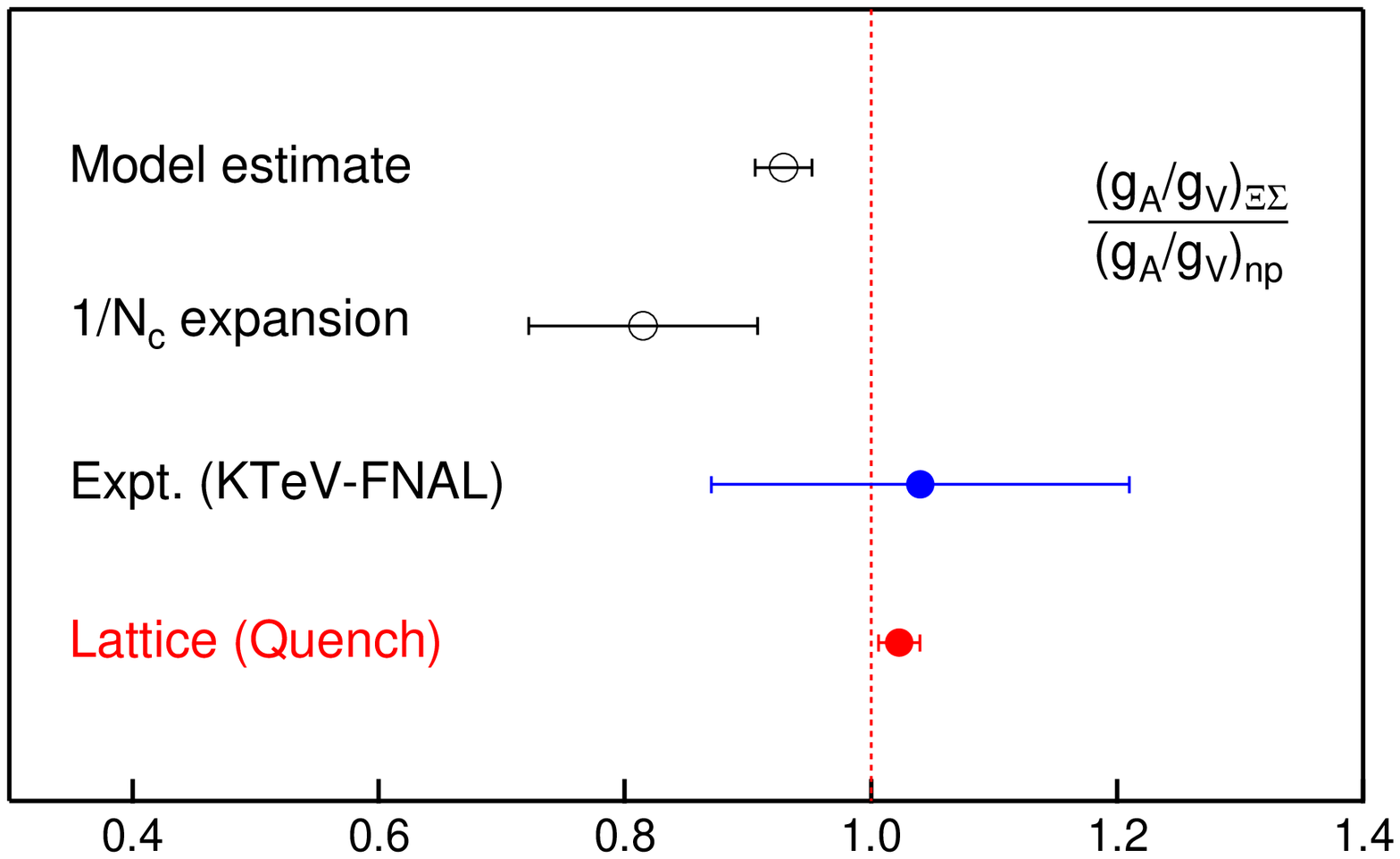}
\caption{Comparison among model predictions, experimental data (KTeV-FNAL)
and our lattice result for the ratio $\frac{(g_A/g_V)_{\Xi\Sigma}}{(g_A/g_V)_{np}}$.}

\label{FIG:g1f1ratioComp}

\ec
\end{minipage}
\end{figure}

We first consider the following double ratio of zero-momentum three-point functions 
($|{\bf q}|^2=|{\bf p}|^2=0$) with fixed $t$:
\be
R(t)=
\frac{
{\rm Tr}[{\cal P}_5^k\langle\Sigma(t')
{\bar u}(t)\gamma_5\gamma_k s(t)
\overline{\Xi}(0)\rangle]
\cdot{\rm Tr}[{\cal P}^4\langle p(t')
{\bar u}(t)\gamma_4 d(t)
\overline{n}(0)\rangle]
}{
{\rm Tr}[{\cal P}^4\langle\Sigma(t')
{\bar u}(t)\gamma_4 s(t)
\overline{\Xi}(0)\rangle]
\cdot{\rm Tr}[{\cal P}_5^k\langle p(t')
{\bar u}(t)\gamma_5\gamma_k d(t)
\overline{n}(0)\rangle]
},
\ee
where projection operators are defined as ${\cal P}^4=\frac{1+\gamma_4}{2}\gamma_4$ and ${\cal P}_{5}^{k}=\frac{1+\gamma_4}{2}\gamma_5\gamma_k$ for the spatial index $k$. 
In this study, the polarization direction $k$ is fixed to be along the $z$-axis. 
All the kinematical factors and  normalization of states are exactly canceled out 
in this ratio. Here, we also take into account the relation between
the lattice renormalizations of the local vector and axial-vector current operators 
as $Z_V=Z_A$ due to the good chiral property of DWFs~\cite{Sasaki:2003jh}. 
As a result, $R(t)$ gives 
\be
R(t)
\xrightarrow[t, (t'-t) \to \infty]{} 
\left.\frac{g_1(q_{\rm max}^2)-\delta g_2(q_{\rm max}^2)}
{f_1(q_{\rm max}^2)+\delta f_3(q_{\rm max}^2)}
\right/
\left(\frac{g_1(0)}{f_1(0)}\right)_{SU(3)} 
=
\frac{g_1(0)/f_1(0)}{\left(g_1(0)/f_1(0)\right)}_{SU(3)}
+ {\cal O}(\delta^2), \\
\label{Eq:DoublRatio}
\ee
where $q_{\rm max}^2=-(\Delta M)^2$, which is proportional to $\delta^2$. 
The $q^2$ dependence of form factors at $q^2=q^2_{\rm max}$ can be involved
in the second-order correction: 
$f_1(q_{\rm max}^2)=f_1(0)+{\cal O}(\delta^2)$
and $g_1(q_{\rm max}^2)=g_1(0)+{\cal O}(\delta^2)$. The non-zero value of the 
second-class form factors $f_3$ and $g_2$ starts from the first-order correction. 
The double ratio $R(t)$ exhibits the first order of the $SU(3)$ breaking effect 
on $g_1(0)/f_1(0)$ since the axial-vector form factors are not protected by
the Ademollo-Gatto theorem. The double ratio $R(t)$ becomes unity in the $SU(3)$ limit.
Therefore, the effect of the first-order $SU(3)$ breaking
on $g_A/g_V=f_1(0)/g_1(0)$ is exposed to this ratio by a deviation from unity.
In other words, the double ratio can be described by $R(t)\rightarrow 1 +\Delta$
with the deviation $\Delta$, which is expected to be proportional to the first-order correction 
of the $SU(3)$ breaking $\delta$ for small $\delta$. 

In Fig.~(\ref{FIG:g1f1ratio}), we plot the deviation from unity in the double ratio 
as a function of the $SU(3)$ breaking parameter $\delta$. Surprisingly, it is observed that
the deviation $\Delta$ is less than a few percents with $\sim 6$ \% $SU(3)$ breaking given 
among octet baryon masses. However, the deviation $\Delta$ is linearly correlated 
with the parameter $\delta$ as we expected. Thus, we may linearly extrapolate 
simulated $\Delta$ to the physical point at $\delta=(M_{\Xi}-M_{\Sigma})/M_{\Xi}=0.0954$. 
We finally obtain a value for the ratio of $(g_A/g_V)_{\Xi\Sigma}$ to $(g_A/g_V)_{np}$
as 1.023(18). Such tiny $SU(3)$ breaking effect on $g_A/g_V$ 
in the $\Xi^0\rightarrow \Sigma^+$ beta decay agrees well with the KTeV experiment
where the value of 1.04(17) is measured.  It is worth mentioning
that our statistical error is at least one order smaller than that of the 
current experiment. In Fig.~(\ref{FIG:g1f1ratioComp}), we summarize our result and
the experimental value combined with predictions from the center-of-mass correction
approach~\cite{Ratcliffe:1998su}
and the $1/N_c$ expansion approach~\cite{Flores-Mendieta:1998ii}. 
A sign of our observed $SU(3)$ breaking correction is opposite to those predictions, 
although the first-order $SU(3)$ breaking on $g_A/g_V$ is accidentally too small 
to justify neglect of the second-order correction in our analysis. 

\subsection{Second-class form factors}

\begin{figure}
\bc
\includegraphics[width=.47\textwidth]{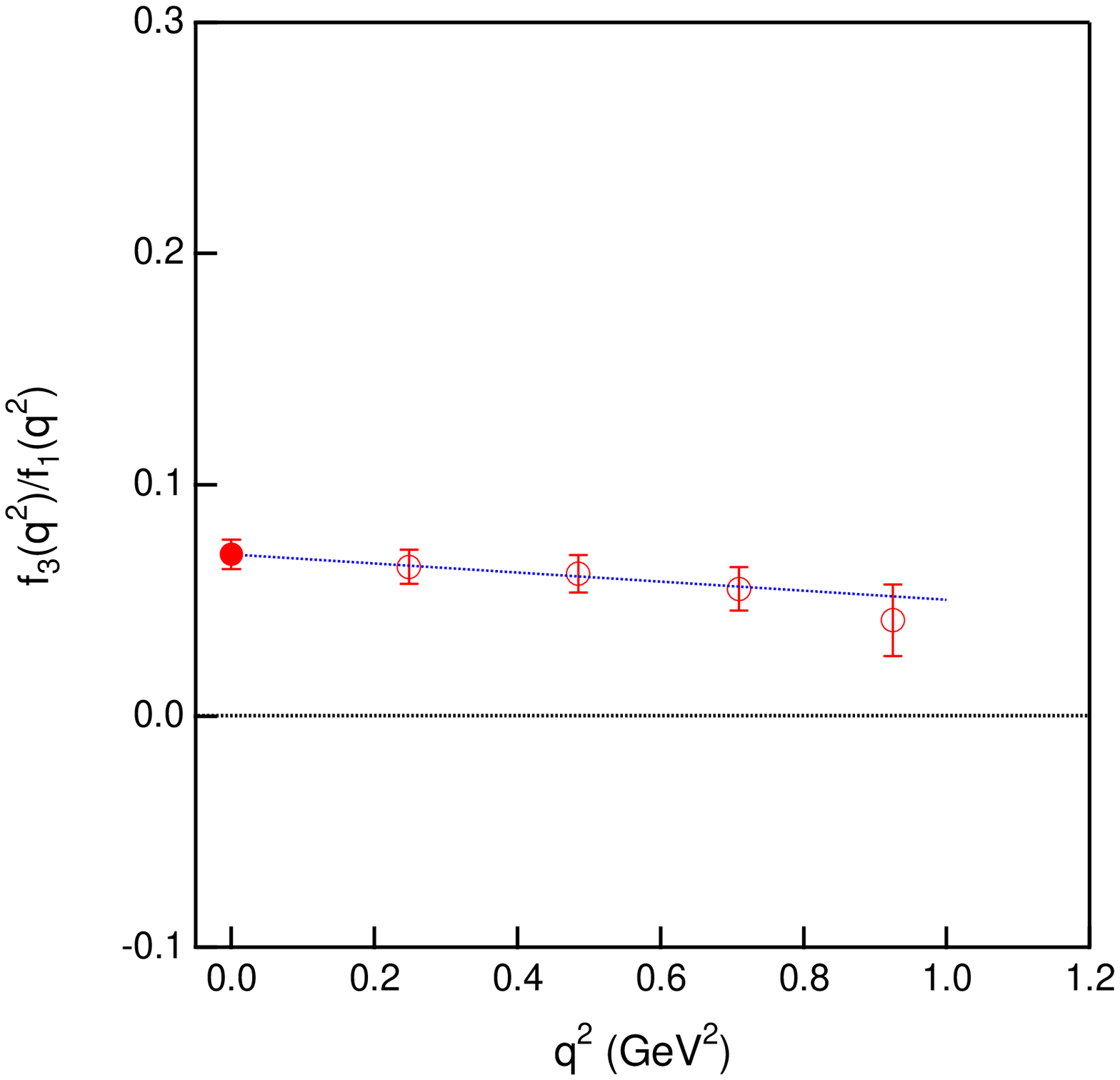}
\includegraphics[width=.47\textwidth]{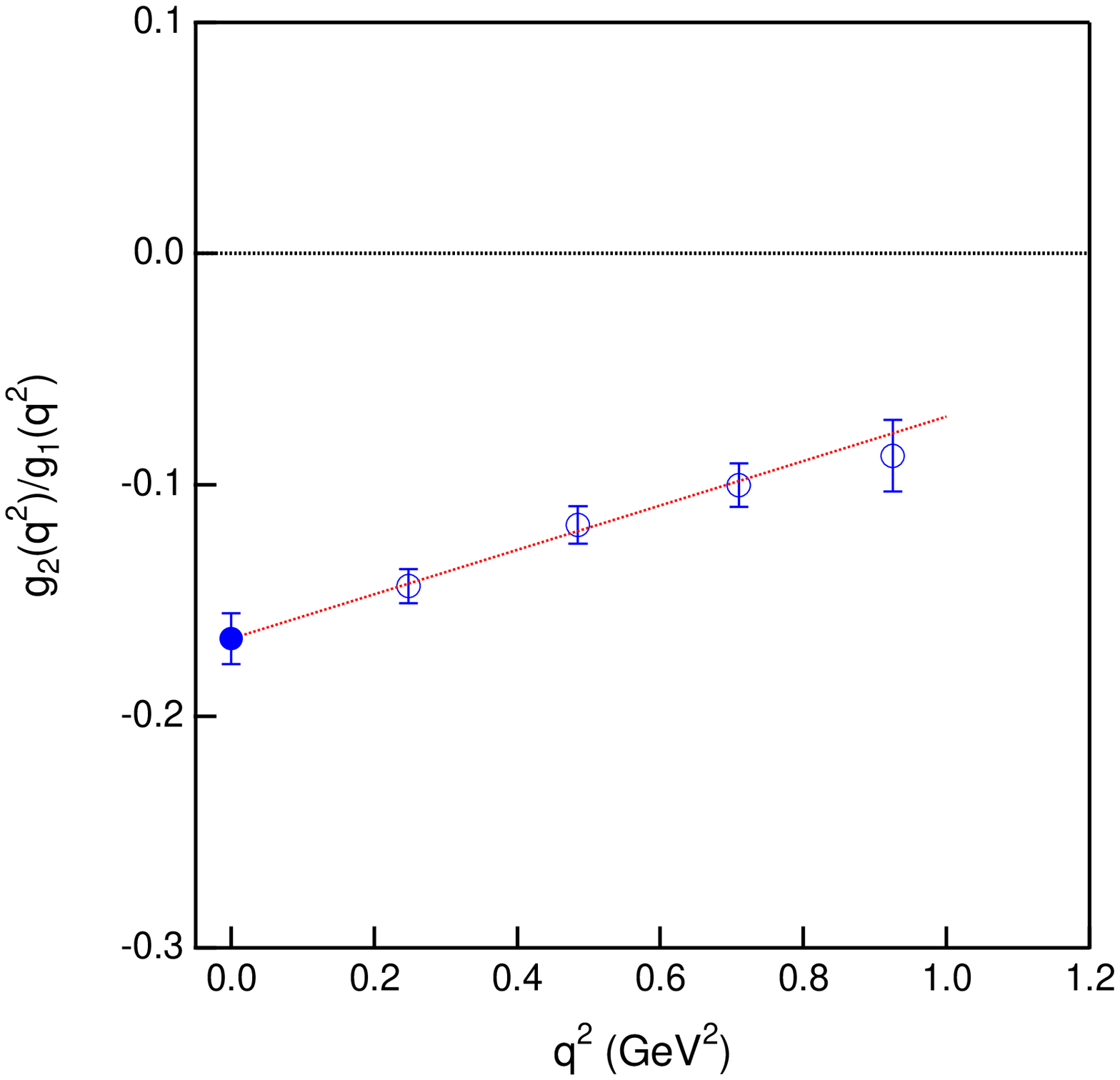}
\caption{The ratios $f_3(q^2)/f_1(q^2)$ (left panel)
and $g_2(q^2)/g_1(q^2)$ (right panel) as functions
of $q^2$ as in the case of $(m_{ud}, m_s)=(0.05, 0.08)$.}

\label{FIG:2ndClass}

\ec
\end{figure}

The kinematics of $|{\bf p}'|^2=|{\bf p}|^2=0$ only allows the particular 
components of the vector (time component) or axial-vector currents (spatial components) 
to access the desired matrix element~\cite{Sasaki:2003jh}. However, in the case if 
either ${\bf p}'$ or ${\bf p}$ are non-zero, three different types of three-point correlation functions are calculated in either channels: 
%
%
\be
\begin{array}{ccc}
\tr \{{\cal P}^4 G_{4}^{V}({\bf p}\neq{\bf 0},{\bf p}'={\bf 0})\},                  &
\tr \{{\cal P}^4 G_{1,2,3}^{V}({\bf p}\neq {\bf 0},{\bf p}'={\bf 0})\},        & 
\tr \{{\cal P}^3_5 G_{1,2}^{V}({\bf p}\neq{\bf 0},{\bf p}'={\bf 0})\}
\end{array}
\ee
for the vector channel, and
%
%
\be
\begin{array}{ccc}
\tr \{{\cal P}^3_5 G_{3}^{A}({\bf p}\neq{\bf 0},{\bf p}'={\bf 0})\},        &
\tr \{{\cal P}^3_5 G_{1,2}^{A}({\bf p}\neq{\bf 0},{\bf p}'={\bf 0})\}, &
\tr \{{\cal P}^3_5 G_{4}^{A}({\bf p}\neq{\bf 0},{\bf p}'={\bf 0})\}
\end{array}
\ee
for the axial-vector channel. As a result, we can solve the simultaneous linear 
equations to get each form factor. In this study, we choose $|{\bf p}'|^2=0$ and 
$|{\bf p}|^2=1,2,3$ and 4 in units of $(2\pi/aL)^2$. 

The second-class form factors $f_3$ and $g_2$ vanish 
in the $SU(3)$ flavor-symmetry limit and non-zero values are induced by the first-order
correction. Although observation of non-zero second-class form factors 
corresponds to the direct signal of the $SU(3)$ breaking effects in hyperon beta decay,
it is hard to measure those form factors in experiment. The KTeV experiment
reported no evidence for a non-zero second-class form factor $g_2$, measuring
$g_2(0)/f_1(0)=-1.7\pm2.0$ which corresponds to $g_2(0)/g_1(0)\simeq-1.4\pm 1.7$.

Figs.~\ref{FIG:2ndClass} show 
ratios of $f_3(q^2)/f_1(q^2)$ (left panel) and $g_2(q^2)/g_1(q^2)$ (right panel) 
for the $\Xi^0 \rightarrow \Sigma^+$ beta decay 
as in the case of $(m_{ud},m_s)=(0.05, 0.08)$ which yields $\delta=0.028(1)$.
We observe non-negligible values of both second-class form factors at simulated $q^2$
in the range of 0.25 to 0.93 ${\rm GeV}^2$.
The simplest linear form is adopted for the $q^2$ extrapolation of form factors
to $q^2=0$ because of mild $q^2$ dependence. Furthermore, we linearly extrapolate measured $f_3(0)/f_1(0)$ and $g_2(0)/g_1(0)$ at five simulated  points of $\delta$ 
to the physical point at $\delta=0.0954$ and then obtain
%
%
\be
\left(\frac{f_3(0)}{f_1(0)}\right)_{\Xi \rightarrow \Sigma}=+0.250(22),
\;\;\;\;\;
\left(\frac{g_2(0)}{g_1(0)}\right)_{\Xi \rightarrow \Sigma}=-0.588(39),
\ee
which show firm evidence for non-zero second-class form factors.

\subsection{Second-order correction on $f_1(0)$}

\begin{figure}
\bc
\includegraphics[width=.47\textwidth]{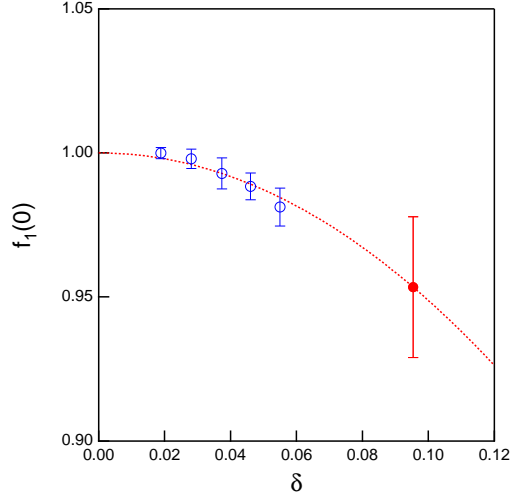}
\caption{
Extrapolation of $f_1(0)$ from simulated values (open circles) to the physical point
(full circle) at $\delta=0.0954$ using the simple fitting form as $f_1(0)=1+ c_2 \delta^2$
suggested by the Ademollo-Gatto theorem.
}

\label{FIG:2ndOrd_on_f1}

\ec
\end{figure}

The value of $f_1(0)$ for the semileptonic hyperon decays
is equal to the $SU(3)$ Clebsh-Gordan coefficient up to the second order of 
the $SU(3)$ breaking, thanks to the Ademollo-Gatto theorem~\cite{Ademollo:1964sr}. 
Here, in the case of the $\Xi^0 \rightarrow \Sigma^+$ process, 
it can be expressed by $f_1(0)=1+{\cal O}(\delta^2)$. 
Although theoretical accurate estimate of $f_1(0)$ 
are highly required for the precise determination of $|V_{us}|$,  
even a sign of the second-order correction, 
is somewhat controversial at present~\cite{Mateu:2005wi}. 

In Fig.~\ref{FIG:2ndOrd_on_f1}, we plot our obtained $f_1(0)$ as a function
of the $SU(3)$ breaking $\delta$. 
The quadratic dependence of the $SU(3)$ breaking appears clearly.
We extrapolate the measured values of $f_1(0)$
to the physical point at $\delta=0.0954$ with a simple fitting form, 
$f_1(0)=1+c_2\delta^2$, according to the Ademollo-Gatto theorem and then obtain
\be
f_1(0)=0.953(24),
\ee
which indicates the negative correction of the second-order breaking
on $f_1(0)$. By combining with a single estimate of $|V_{us}f_1(0)|=0.216(33)$
from the KTeV experiment~\cite{Alavi-Harati:2001xk}, we finally obtain
\be
|V_{us}|=0.219(27)_{\rm exp}(5)_{\rm theory},
\ee
which is consistent with the value obtained from $K_{l3}$ decays
and the CKM unitary predicted value~\cite{Dawson:2005zv}.

\section{Summary}

We have presented preliminary results of the flavor $SU(3)$ breaking effect
in the $\Xi^0 \rightarrow \Sigma^+$ beta decay using quenched DWF simulations.
We found a tiny first-order correction on $g_A/g_V$ in agreement with the
KTeV experiment. Although our result doesn't conflict with Cabibbo model fits~\cite{Cabibbo:2003cu}, it doesn't mean that there is no $SU(3)$ breaking effect in hyperon beta decay. 
Indeed, we measured the non-zero value of the second-class form factors $f_3$ and $g_2$,
which correspond to the direct evidence of the $SU(3)$ breaking effect. 
We finally observed that the second-order correction on $f_1(0)$ is negative. 
This leads to the closer value of $|V_{us}|$ to the value obtained from $K_{l3}$ decays,
while our observed tendency for the $SU(3)$ breaking correction is opposite against 
predictions of both HBChPT and large $N_c$ analysis.


\end{document}